\documentclass[aps,pre,reprint,groupedaddress]{revtex4-2}

\usepackage{amsmath,amssymb} 
\usepackage{graphicx}

\usepackage{bm}
\usepackage{booktabs}

\begin{document}


\title{Memory Uncertainty Relation and Harmonic Memory in Random Recurrent Networks}


\author{Taichi Haruna}
\email[]{tharuna@lab.twcu.ac.jp}
\affiliation{Department of Information and Mathematical Sciences, School of Arts and Sciences, Tokyo Woman's Christian University, 2-6-1 Zempukuji, Suginami-ku, Tokyo 167-8585, Japan}

\author{Kohei Nakajima}
\email[]{k-nakajima@isi.imi.i.u-tokyo.ac.jp}
\affiliation{Graduate School of Information Science and Technology, The University of Tokyo, Bunkyo-ku, Tokyo 113-8656, Japan}


\date{\today}

\begin{abstract}
We present an inequality that bounds the short-term memory capability of dynamical systems from below. It can be interpreted as an uncertainty relation between a measure of short-term memory and that of the size of state fluctuations induced by input signals. The lower bound can be achieved by a readout weight and thus represents a suboptimal memory called harmonic memory. We examine analytically and numerically the inequality in a number of reservoir systems subject to input noise. We illustrate cases in which equality is achieved exactly, equality holds asymptotically, and the inequality is strict. We also study the effect of a state-space regularization to elucidate the inequality in terms of the fluctuation structure of the state-space. We find that a certain strength of input noise induces extra memory under the regularization, and we refer to this phenomenon as noise-induced memory. We observe that the memory uncertainty relation does not hold in general for the regularized memory and harmonic memory. This fact is explained in terms of the mechanism of noise-induced memory. 
\end{abstract}


\maketitle

\section{Introduction}
\label{sec:intro}
A prerequisite for a dynamical system to perform temporal information processing is that it has short-term memory capability. Namely, to emit a desired output signal depending on a recent sequence of temporarily varying input signals, the system must somehow store the input signals within its state and be able to reconstruct them. Studies of reservoir computing have shown that random recurrent neural networks and even physical systems can achieve this characteristic by utilizing their transient dynamics induced by input signals~\cite{Maass2002,Jaeger2004,Lukosevicius2009,Tanaka2019,Nakajima2020,Nakajima2021}. 

The short-term memory capability of a dynamical system can be quantified by measuring how the current state of the system is correlated with or dependent on past input signals. Building on this idea, several measures of short-term memory have been proposed such as memory capacity~\cite{Jaeger2002}, information processing capacity~\cite{Dambre2012}, and Fisher memory~\cite{Ganguli2008}. These measures have been extensively studied from different perspectives: analytic expressions of memory capacity~\cite{White2004,Hermans2010} and Fisher memory~\cite{Tino2018} in linear random recurrent neural networks, mean-field calculations in nonlinear cases~\cite{Toyoizumi2011,Schuecker2018,Haruna2019}, optimality in relation to the edge of chaos~\cite{Bertschinger2004,Boedecker2012,Farkas2016}, effects of correlations~\cite{Marzen2017,Gonon2020} or noise~\cite{Guan2025} in input signals, and memory-nonlinearity trade-off relations~\cite{Dambre2012,Inubushi2017,Kubota2021,SchultetoBrinke2023}. For linear random recurrent neural networks, the structure of the state-space related to short-term memory has also been studied in terms of feature space representations of input signals using temporal kernels~\cite{Tino2020} and encoding of input signals by controllability matrices~\cite{Verzelli2021}. 

Here, we investigate the relation between the statistical structure of the state-space and the memory function that yields the memory capacity when summed over its argument $k$ representing the time delay of the input signals. In particular, we give a simple inequality that bounds the memory function from below and thus can be regarded as an uncertainty relation. The lower bound is the reciprocal of the relative size of state fluctuations in the average direction where the past input signal entered in the system $k$ steps before is stored. We show that the lower bound can be achieved by a specific readout weight vector, namely, it represents a suboptimal memory. We call it \textit{harmonic memory}. In the following, we restrict our focus on a class of random recurrent networks subject to input noise to obtain analytical results, although the inequality holds for broader class of dynamical systems. We also examine the relation between the memory and harmonic memory capacities under a state-space regularization called \textit{principal component regularization (PCR)} to shed light on the role of the direction in the state-space on which the harmonic memory is realized in relation to the principal component structure of the state fluctuations. We find that PCR gives rise to \textit{noise-induced memory (NIM)} in which a certain level of noise in input signals enhances the short-term memory capability. We show that the degree of the enhancement can be different between the regularized memory and harmonic memory capacities, and even the latter can become larger than the former for a certain reservoir network architecture. 

This paper is organized as follows. In Sec.~\ref{sec:model}, we introduce the recurrent neural network model that we use in this paper and review the memory function associated with it. In Sec.~\ref{sec:results}, first we derive the lower bound of the memory function and show that it represents a suboptimal memory. The inequality is studied through examples with different reservoir network architectures. Second, we introduce PCR and explain the mechanism of NIM and its effect on the relation between the regularized memory and harmonic memory function as described above. Finally, in Sec.~\ref{sec:conclusion}, we give several remarks on connections between our results and the existing literature. 

\section{Model}
\label{sec:model}
Specifically, we consider the following discrete-time recurrent neural network model called an echo-state network (ESN) with $N$ neurons~\cite{Jaeger2001,White2004,Massar2013}. Let $x_n(t) \in \mathbb{R}$ be the state of neuron $n$ ($1 \leq n \leq N$) in the reservoir layer at time step $t$. We write $\bm{x}(t)=(x_1(t),x_2(t),\dots,x_N(t))^\mathrm{T}$, where $\mathrm{T}$ represents the transpose of a vector or a matrix. The time evolution of $\bm{x}(t)$ obeys the following equation: 
\begin{equation}
\bm{x}(t+1) = \bm{f}\left( W \bm{x}(t) + s(t+1)\bm{v} + \bm{z}(t) \right),
\label{eq:esn}
\end{equation}
where $s(t+1)$ is an input signal given at time step $t+1$, $\bm{v}=(v_1,v_2,\dots,v_N)^\mathrm{T} \in \mathbb{R}^N$ is a weight vector between the input and the reservoir neurons, $W=\left( w_{mn} \right)_{1 \leq m,n \leq N}$ is a connection matrix of the reservoir layer with $w_{mn} \in \mathbb{R}$ a weight from neuron $n$ to neuron $m$, $\bm{z}(t)=(z_1(t),z_2(t),\dots,z_N(t))^\mathrm{T} \in \mathbb{R}^N$ is a noise vector, and $\bm{f}=(f_1,f_2,\dots,f_N)$ with $f_n$ an activation function of neuron $n$. The output signal $y(t)$ is the state of the readout neuron at time step $t$ that is given by 
\begin{equation}
y(t) = \sum_{i=1}^L u_i x_i(t),
\label{eq:output}
\end{equation}
where we assume that the indices of the reservoir neurons connected to the readout neuron are $i=1,2,\dots,L$ ($L \leq N$) without loss of generality. We write $\bm{u}=(u_1,u_2,\dots,u_L) \in \mathbb{R}^L$. By abuse of notation, we write $\bm{x}(t)=(x_1(t),x_2(t),\dots,x_L(t))^\mathrm{T} \in \mathbb{R}^L$. Thus, $y(t)=\bm{u}^\mathrm{T}\bm{x}(t)$. Given a target output signal $d(t)$, the readout weight vector $\bm{u}$ is optimized under the condition that the mean squared error $\langle \left| y(t) - d(t) \right|^2 \rangle_t$ is minimized, where $\langle \dots \rangle_t$ is the time average. 

Let $s(t)$ be a stationary input signal with mean $0$ and variance $1$. For each $k=0,1,2,\dots$, we consider $d(t)=s(t-k)$ as the target signal. The readout weight vector $\bm{u}$ that minimizes $\langle \left| y(t) - s(t-k) \right|^2 \rangle_t$ is given by $\bm{u} = \bm{u}_k = C^{-1}\bm{p}_k$, where 
\begin{equation}
C=\langle \bm{x}(t)\bm{x}(t)^\mathrm{T} \rangle_t 
\label{eq:c}
\end{equation}
is an $L \times L$ symmetric matrix and $\bm{p}_k \in \mathbb{R}^L$ is defined by 
\begin{equation}
\bm{p}_k = \langle s(t-k)\bm{x}(t) \rangle_t. 
\label{eq:pk}
\end{equation}
We assume that the time averages defining $C$ and $\bm{p}_k$ exist, and that $C$ is positive definite so that $C^{-1}$ exists. We put \begin{equation}
\hat{y}_k(t)=\bm{u}_k^\mathrm{T}\bm{x}(t) = \bm{p}_k^\mathrm{T}C^{-1}\bm{x}(t). 
\label{eq:hatyk}
\end{equation}
The coefficient of determination defined as 
\begin{align}
m(k) &= 1 - \langle \left| \hat{y}_k(t) - s(t-k) \right|^2 \rangle_t \nonumber\\
&= 1 - \langle \hat{y}_k(t)^2 \rangle_t + 2 \langle \hat{y}_k(t)s(t-k) \rangle_t - \langle s(t-k)^2 \rangle_t \nonumber\\
&= - \bm{p}_k^\mathrm{T}C^{-1}\bm{p}_k + 2\bm{p}_k^\mathrm{T}C^{-1}\bm{p}_k \nonumber\\
&= \bm{p}_k^\mathrm{T}C^{-1}\bm{p}_k
\label{eq:mk}
\end{align}
is a measure of the degree of reconstruction of $s(t-k)$ from $\hat{y}_k(t)$ for each $k=0,1,2,\dots$, where we have used $\langle s(t-k)^2 \rangle_t = 1$ and $\langle \hat{y}_k(t)^2 \rangle_t = \bm{p}_k^\mathrm{T}C^{-1}\bm{p}_k$ to obtain the third line. The latter holds since $\langle \hat{y}_k(t)^2 \rangle_t = \bm{p}_k^\mathrm{T}C^{-1}\langle \bm{x}(t)\bm{x}(t)^\mathrm{T} \rangle_t C^{-1}\bm{p}_k = \bm{p}_k^\mathrm{T}C^{-1} C C^{-1}\bm{p}_k = \bm{p}_k^\mathrm{T}C^{-1}\bm{p}_k$. We note that this gives an alternative expression of $m(k)$, namely, $m(k)=\langle \hat{y}_k(t)^2 \rangle_t$. $m(k)$ is called the memory function and satisfies $0 \leq m(k) \leq 1$~\cite{Jaeger2002}. 

\section{Results}
\label{sec:results}

\subsection{Memory Uncertainty Relation and Harmonic Memory}
\label{subsec:murhm}
We introduce a quantity 
\begin{equation}
h(k)=\frac{\| \bm{p}_k \|^2}{\overline{\bm{p}}_k^\mathrm{T}C\overline{\bm{p}}_k}\label{eq:hk}
\end{equation}
when $\| \bm{p}_k \| > 0$ and otherwise $h(k)=0$ for each $k=0,1,2,\dots$, where $\overline{\bm{p}}_k=\frac{\bm{p}_k}{\| \bm{p}_k \|}$ and $\| \bm{p}_k \|=\sqrt{\bm{p}_k^\mathrm{T}\bm{p}_k}$. We show that $h(k)$ is a lower bound of $m(k)$, namely, the following inequality holds: 
\begin{equation}
m(k) \geq h(k).
\label{eq:mkhk}
\end{equation}
The condition of equality is described later.

To see that Eq.~(\ref{eq:mkhk}) holds, let $\mu_1 \geq \mu_2 \geq \dots \geq \mu_L > 0$ be the eigenvalues of the positive definite matrix $C$, with $\bm{q}_i$ the eigenvector corresponding to $\mu_i$ ($i=1,2,\dots,L$). We can take the eigenvectors $\bm{q}_1,\bm{q}_2,\dots,\bm{q}_L$ so that they form an orthonormal basis of $\mathbb{R}^L$. Thus, 
\begin{equation}
C=\sum_{i=1}^L \mu_i \bm{q}_i \bm{q}_i^\mathrm{T}. 
\label{eq:cexpansion}
\end{equation}
It is enough to consider the case $\| \bm{p}_k \| > 0$. Substituting 
\begin{equation}
C^{-1} = \sum_{i=1}^L \mu_i^{-1} \bm{q}_i \bm{q}_i^\mathrm{T} 
\label{eq:c-1}
\end{equation}
into Eq.~(\ref{eq:mk}) and using Jensen's inequality, we have 
\begin{equation}
m(k) = \| \bm{p}_k \|^2 \sum_{i=1}^L \mu_i^{-1} \pi_{k,i} 
\geq \frac{\| \bm{p}_k \|^2}{\sum_{i=1}^L \mu_i \pi_{k,i}} 
= h(k), 
\label{eq:proof}
\end{equation}
where we have used the fact that the nonnegative quantities 
\begin{equation}
\pi_{k,i} = \left( \overline{\bm{p}}_k^\mathrm{T}\bm{q}_i \right)^2 = \frac{\left( \bm{p}_k^\mathrm{T}\bm{q}_i \right)^2}{\| \bm{p}_k \|^2}
\label{eq:piki}
\end{equation}
($i=1,2,\dots,L$) form a probability distribution $\bm{\pi}_k=(\pi_{k,1},\pi_{k,2},\dots,\pi_{k,L})$ on the set $\{1,2,\dots,L\}$. The equality in Eq.~(\ref{eq:mkhk}) holds if and only if $\mu_i$ is constant on the support of $\bm{\pi}_k$. 

Equation~(\ref{eq:mkhk}) can be interpreted as an uncertainty relation analogous to the thermodynamic uncertainty relation that gives a trade-off relation between the entropy production rate and the relative size of fluctuations of the probability currents in a wide range of nonequilibrium thermodynamic systems~\cite{Barato2015,Gingrich2016}, although the derivation of Eq.~(\ref{eq:mkhk}) is much simpler. For each $k=0,1,2,\dots$, $f(k)=h(k)^{-1}$ is the relative size of the second moment of $\bm{x}(t) \in \mathbb{R}^L$ in the direction of $\bm{p}_k$. In particular, if $\langle \bm{x}(t) \rangle_t = \bm{0}$, then $f(k)$ is regarded as the relative size of the variance of $\bm{x}(t)$ in the direction of $\bm{p}_k$. If we write Eq.~(\ref{eq:mkhk}) as $m(k)f(k) \geq 1$, then this represents a trade-off relation between $m(k)$ and $f(k)$: One cannot make both quantities small simultaneously. We also note that it is an instance of the Cram\'{e}r--Rao inequality~\cite{Cover2006}: $f(k) \geq m(k)^{-1}$. Indeed, if we consider the Gaussian probability density function $p(\bm{x}; \theta) \propto \exp\left( -\frac{1}{2} (\bm{x} - \theta \bm{p}_k)^\mathrm{T}C^{-1}(\bm{x} - \theta \bm{p}_k) \right)$ on $\mathbb{R}^L$ with the parameter $\theta \in \mathbb{R}$, then $\hat{\theta}(\bm{x})=\frac{\bm{p}_k^\mathrm{T}\bm{x}}{\| \bm{p}_k \|^2}$ is an unbiased estimator of $\theta$. It is straightforward to see that $f(k)$ is the variance of $\hat{\theta}$ and $m(k)$ the Fisher information for $p(\bm{x}; \theta)$. 

Alternatively, Eq.~(\ref{eq:mkhk}) can be seen as an instance of the AM-HM inequality~\cite{Hardy1988}, namely, the arithmetic mean of given positive real numbers is greater than or equal to their harmonic mean. Indeed, $m(k)=\sum_{i=1}^L m_i(k) \pi_{k,i}$ is the arithmetic mean of 
\begin{equation}
m_i(k) = \frac{\| \bm{p}_k \|^2}{\mu_i}. 
\label{eq:mik}
\end{equation}
However, $h(k)$ is the harmonic mean of $m_i(k)$. Hence, Eq.~(\ref{eq:mkhk}) follows. 

Next, we show that $h(k)$ is not only a lower bound of $m(k)$, but also represents a memory. Namely, it is a coefficient of determination corresponding to a certain readout weight $\bm{u}$. Indeed, for an arbitrary readout weight $\bm{u} \in \mathbb{R}^L$, the coefficient of determination is given by 
\begin{align}
1 - \langle \left| y(t) - s(t-k) \right|^2 \rangle_t &= 2 \langle s(t-k)y(t) \rangle_t - \langle y(t)^2 \rangle_t \nonumber\\
&= 2\bm{u}^\mathrm{T}\bm{p}_k - \bm{u}^\mathrm{T}C\bm{u}.
\label{eq:dethk}
\end{align}
Let us restrict $\bm{u}$ in the direction of $\bm{p}_k$ and write $\bm{u}=\sigma \bm{p}_k$ ($\sigma \in \mathbb{R}$). It is straightforward to show that Eq.~(\ref{eq:dethk}) is maximized when $\sigma=\sigma_k$ and the corresponding maximum value is $h(k)$, where 
\begin{equation}
\sigma_k = \frac{1}{\overline{\bm{p}}_k^\mathrm{T}C\overline{\bm{p}}_k}. 
\label{eq:sigk}
\end{equation}
Thus, $h(k)$ is a suboptimal memory realized using a specific readout weight, and we call it the \textit{harmonic memory function}. We write 
\begin{equation}
\hat{y'}_k(t) = \sigma_k \bm{p}_k^\mathrm{T}\bm{x}(t), 
\label{eq:hatypk}
\end{equation}
and we have $h(k) = \langle \hat{y'}_k(t)^2 \rangle_t$. Indeed, $\langle \hat{y'}_k(t)^2 \rangle_t = \sigma_k^2 \bm{p}_k^\mathrm{T} \langle \bm{x}(t)\bm{x}(t)^\mathrm{T} \rangle_t \bm{p}_k = \sigma_k^2 \bm{p}_k^\mathrm{T} C \bm{p}_k = \sigma_k \| \bm{p}_k \|^2 = h(k)$. 

\begin{figure*}[t]
\centering
\includegraphics[width=16cm,bb=0 0 508 172]{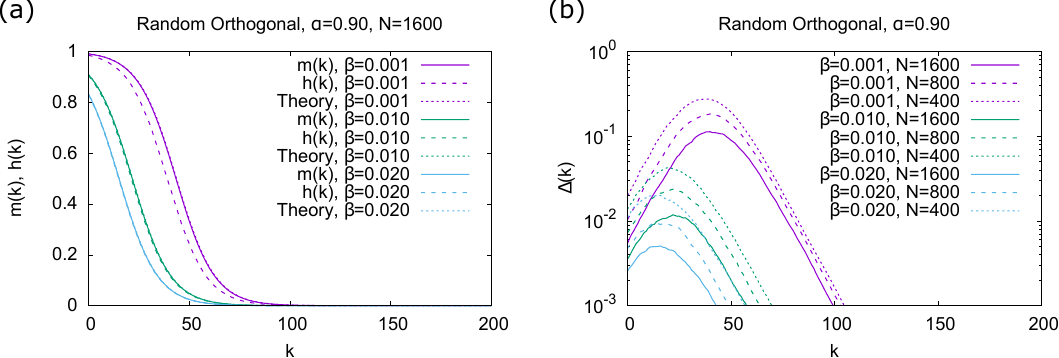}
\caption{
(a) $m(k)$ (solid lines) and $h(k)$ (broken lines) for RONs with $N=1600$, $\alpha=0.9$. The dotted lines are the mean-field curves of Eq.~(\ref{eq:ronmkhk}). (b) $\Delta(k)$ for RONs with $\alpha = 0.9$. For each condition, $m(k)$ and $h(k)$ are averaged over $100$ realizations of RONs. 
}
\label{fig:1}
\end{figure*}

\begin{figure*}[t]
\centering
\includegraphics[width=16cm,bb=0 0 508 172]{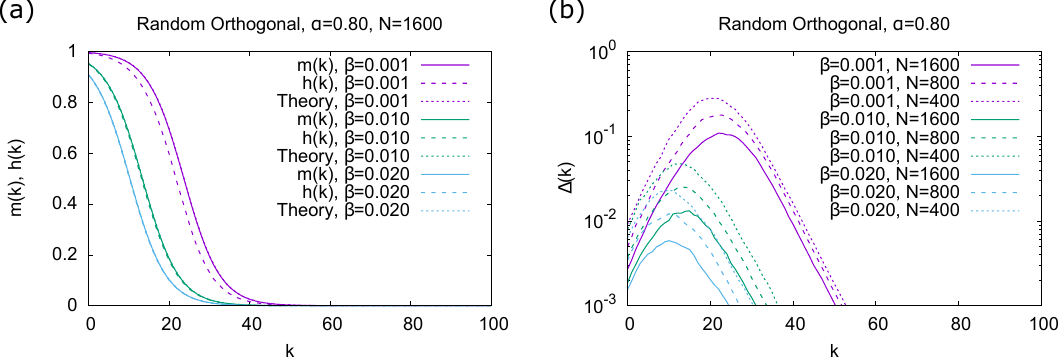}
\caption{
(a) $m(k)$ (solid lines) and $h(k)$ (broken lines) for RONs with $N=1600$, $\alpha=0.8$. The dotted lines are the mean-field curves of Eq.~(\ref{eq:ronmkhk}). (b) $\Delta(k)$ for RONs with $\alpha = 0.8$. For each condition, $m(k)$ and $h(k)$ are averaged over $100$ realizations of RONs. 
}
\label{fig:2}
\end{figure*}

Note that we can think of the state $\bm{x}(t)$ being projected in the direction of $\bm{p}_k$ instead of restricting the readout weight $\bm{u}$ in the direction of $\bm{p}_k$ as just above. Namely, we consider $y(t)=\sigma \bm{p}_k^\mathrm{T}\bm{x}(t)$ and then find $\sigma$ that maximizes the left-hand side of Eq.~(\ref{eq:dethk}). 

We also note that the difference between $m(k)$ and $h(k)$ denoted by $\Delta(k) = m(k) - h(k) \geq 0$ can be expressed as 
\begin{equation}
\Delta(k) = \frac{1}{h(k)}\sum_{i=1}^L \left( m_i(k) - h(k) \right)^2 \rho_{k,i}, 
\label{eq:deltak}
\end{equation}
where 
\begin{equation}
\rho_{k,i} = \frac{\pi_{k,i}}{m_i(k)}h(k) 
\label{eq:rhoki}
\end{equation}
($i=1,2,\dots,L$) form a probability distribution on the set $\{1,2,\dots,L\}$ (Appendix~\ref{sec:a1}). Thus, $\Delta(k)$ is the relative variance of $m_i(k)$ ($i=1,2,\dots,L$) with respect to their harmonic mean such that the individual deviations are weighted by the corresponding contributing factors $\frac{\pi_{k,i}}{m_i(k)}$ to $h(k)$.

\begin{figure*}[t]
\centering
\includegraphics[width=16cm,bb=0 0 508 172]{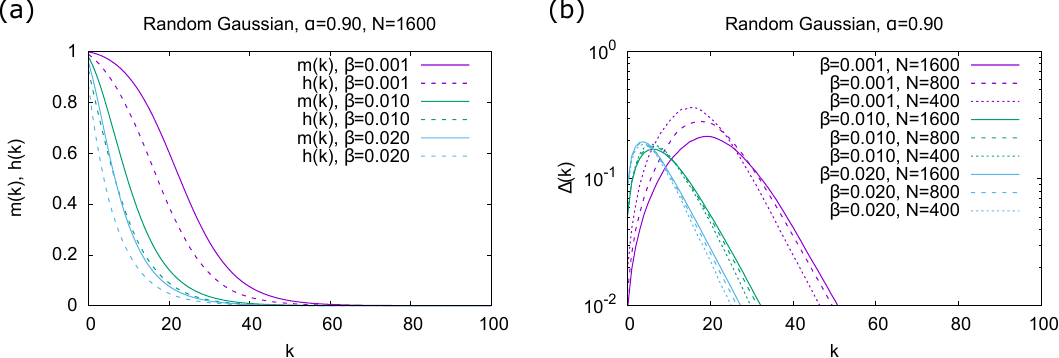}
\caption{
(a) $m(k)$ (solid lines) and $h(k)$ (broken lines) for RGNs with $N=1600$, $\alpha=0.9$. (b) $\Delta(k)$ for RGNs with $\alpha = 0.9$. For each condition, $m(k)$ and $h(k)$ are averaged over $100$ realizations of RGNs. 
}
\label{fig:3}
\end{figure*}

\begin{figure*}[t]
\centering
\includegraphics[width=16cm,bb=0 0 508 172]{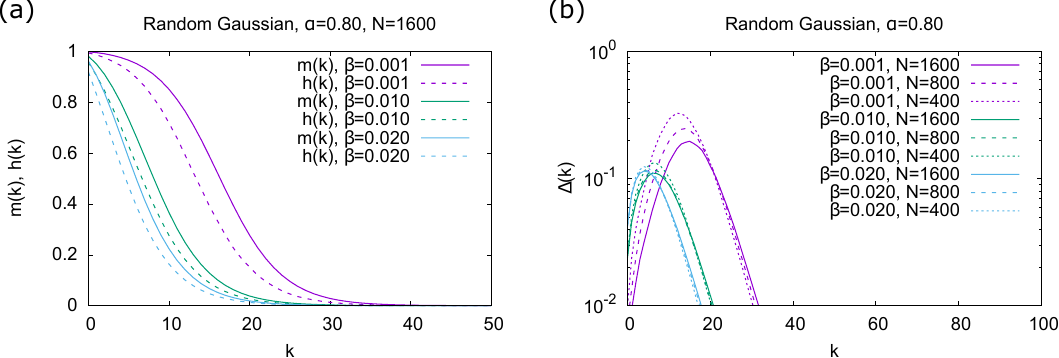}
\caption{
(a) $m(k)$ (solid lines) and $h(k)$ (broken lines) for RGNs with $N=1600$, $\alpha=0.8$. (b) $\Delta(k)$ for RGNs with $\alpha = 0.8$. For each condition, $m(k)$ and $h(k)$ are averaged over $100$ realizations of RGNs. 
}
\label{fig:4}
\end{figure*}

Next, we investigate the behavior of $h(k)$ and its relation to $m(k)$ through examples. In the following examples, we assume that the input signal $s(t)$ satisfies $\langle s(t) \rangle_t=0$ and $\langle s(t)s(t+k) \rangle_t = \iota \delta_{k,0}$ ($k=0,1,2,\dots$), the input weight vector $\bm{v}$ is sampled from the uniform distribution on the unit sphere in $\mathbb{R}^N$, and the noise vector $\bm{z}(t)$ satisfies $\langle \bm{z}(t) \rangle_t = \bm{0}$ and $\langle z_m(t)z_n(t+k) \rangle_t = \beta \delta_{k,0} \delta_{m,n}$ ($1 \leq m \neq n \leq N$, $k=0,1,2,\dots$), where $\delta_{i,j}$ is the Kronecker delta, $\iota > 0$ is the strength of the input signal, and $\beta \geq 0$ is the strength of the noise. The next three examples are linear ($\bm{f}=\mathrm{id}_{\mathbb{R}^N}$) and all the reservoir neurons are connected to the readout ($L=N$). In these cases, we can assume that $\iota = 1$ without loss of generality and we have 
\begin{equation}
\bm{p}_k=W^k \bm{v} 
\label{eq:pklin}
\end{equation}
and 
\begin{equation}
C = C_0 + \beta C_n, 
\label{eq:clin}
\end{equation}
where $C_0 = \sum_{k=0}^\infty \bm{p}_k \bm{p}_k^\mathrm{T}$ and $C_n = \sum_{k=0}^\infty W^k W^{k\mathrm{T}}$~\cite{White2004}. As the fourth example, we consider briefly a nonlinear case with $L=\mathcal{O}(1)$. 

The first example involves distributed shift register networks (DSRNs)~\cite{White2004}. For a DSRN, its connection matrix $W$ is given as follows. Putting $\bm{v}_1=\bm{v}$, we take $\bm{v}_2,\bm{v}_3,\dots,\bm{v}_N \in \mathbb{R}^N$ so that $\bm{v}_1,\bm{v}_2,\dots,\bm{v}_N$ form an orthonormal basis of $\mathbb{R}^N$, and we define $W=\sqrt{\alpha} \sum_{n=1}^{N-1} \bm{v}_{n+1} \bm{v}_n^\mathrm{T}$, where $0 < \alpha < 1$. Since $W\bm{v}_n=\sqrt{\alpha}\bm{v}_{n+1}$ for $1 \leq n \leq N-1$ and $W\bm{v}_N=\bm{0}$, we have $\bm{p}_k = \alpha^\frac{k}{2}\bm{v}_{k+1}$ for $k=0,1,\dots,N-1$ and $\bm{p}_k = \bm{0}$ for $k \geq N$. We also have $C=\sum_{n=1}^N \mu_n \bm{q}_n \bm{q}_n^\mathrm{T}$, with $\mu_n = \alpha^{n-1} + \tilde{\beta}(1-\alpha^n)$ and $\bm{q}_n=\bm{v}_n$ for $n=1,2,\dots,N$, where $\tilde{\beta}=\beta/(1-\alpha)$. Thus, 
\begin{equation}
m(k)=\frac{\alpha^k}{\alpha^k+\tilde{\beta}(1-\alpha^{k+1})}=h(k)
\label{eq:mkhkdsrn}
\end{equation}
for $k=0,1,\dots,N-1$ and $m(k)=h(k)=0$ for $k \geq N$. Indeed, for each $k=0,1,\dots,N-1$, the condition of equality for Eq.~(\ref{eq:mkhk}) holds, since $\pi_{k,n} = \left( \overline{\bm{p}}_k^\mathrm{T}\bm{q}_n \right)^2=\left( \bm{v}_{k+1}^\mathrm{T}\bm{v}_n \right)^2=\delta_{k+1,n}$. 

The second example involves random orthogonal networks (RONs)~\cite{White2004} with $W=\sqrt{\alpha}Q$, where $0 < \alpha < 1$ and $Q$ is an $N \times N$ orthogonal matrix sampled uniformly at random from the set of all $N \times N$ orthogonal matrices. We have 
\begin{equation}
\bm{p}_k = \alpha^\frac{k}{2} \tilde{\bm{v}}_k 
\label{eq:pkron}
\end{equation}
and 
\begin{equation}
C = \sum_{k=0}^\infty \alpha^k \tilde{\bm{v}}_k \tilde{\bm{v}}_k^\mathrm{T} + \tilde{\beta}I, 
\label{eq:cron}
\end{equation}
where 
\begin{equation}
\tilde{\bm{v}}_k = Q^k \bm{v} 
\label{eq:tildevk}
\end{equation}
and $I$ is the identity matrix. For finite $N$, the equality in Eq.~(\ref{eq:mkhk}) does not hold in general, but we find 
\begin{equation}
m(k)=\frac{\alpha^k}{\alpha^k+\tilde{\beta}}=h(k)
\label{eq:ronmkhk}
\end{equation}
for $k=0,1,2,\dots$ as $N \to \infty$ under the annealed approximation such that $Q$ and $\bm{v}$ are resampled at each time step. Since the left-hand equality is shown in \cite{White2004}, here we derive only the right-hand equality (Appendix~\ref{sec:a2}). 

Figure~\ref{fig:1} shows $m(k)$, $h(k)$, and Eq.~(\ref{eq:ronmkhk}) for RONs with $\alpha = 0.9$ and $\beta=0.001, 0.01, 0.02$. For $N=1600$, $m(k)$ and $h(k)$ are approximated well by Eq.~(\ref{eq:ronmkhk}) [Fig.~\ref{fig:1}(a)]. The larger the value of $\beta$, the better the approximation. As $N$ increases, $\Delta(k)$ decreases [Fig.~\ref{fig:1}(b)]. Similar results are obtained for $\alpha=0.8$ (Fig.~\ref{fig:2}). 

\begin{figure*}[t]
\centering
\includegraphics[width=16cm,bb=0 0 561 126]{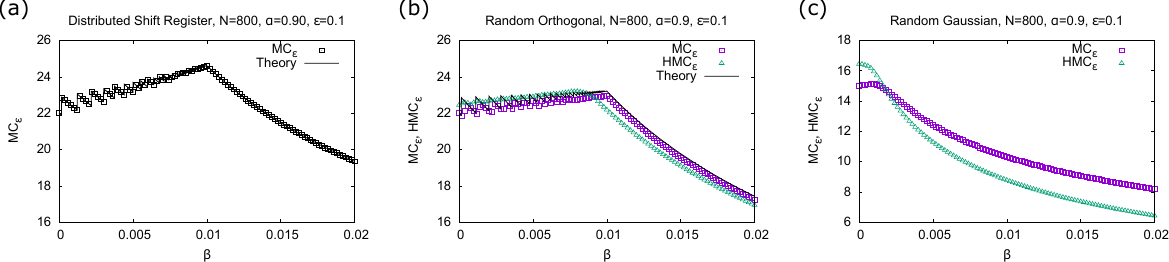}
\caption{
$MC_\epsilon$ and $HMC_\epsilon$ for the three linear reservoirs with $N=800$, $\alpha=0.9$, and $\epsilon=0.1$. (a) DSRN. A single realization of a DSRN is used to compute $MC_\epsilon (= HMC_\epsilon)$ numerically. The solid line is Eq.~(\ref{eq:mcehcedsrn}). (b) RON. The solid line is Eq.~(\ref{eq:hceron}). (c) RGN. In panels (b) and (c), $MC_\epsilon$ and $HMC_\epsilon$ are averaged over $100$ realizations of each linear reservoir. 
}
\label{fig:5}
\end{figure*}

\begin{figure*}[t]
\centering
\includegraphics[width=16cm,bb=0 0 561 126]{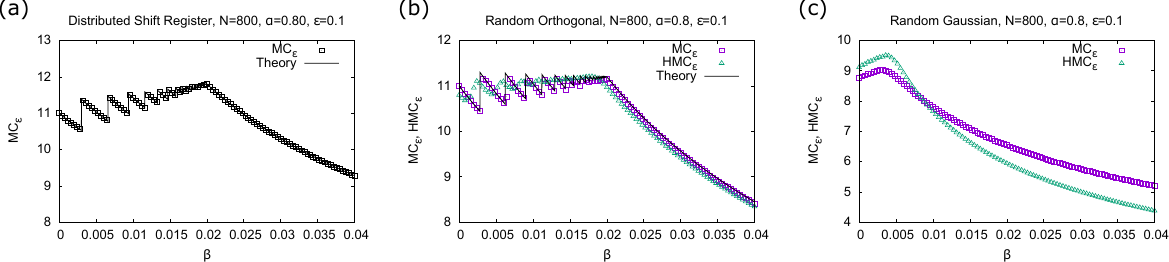}
\caption{
$MC_\epsilon$ and $HMC_\epsilon$ for the three linear reservoirs with $N=800$, $\alpha=0.8$, and $\epsilon=0.1$. (a) DSRN. A single realization of a DSRN is used to compute $MC_\epsilon (= HMC_\epsilon)$ numerically. The solid line is Eq.~(\ref{eq:mcehcedsrn}). (b) RON. The solid line is Eq.~(\ref{eq:hceron}). (c) RGN. In panels (b) and (c), $MC_\epsilon$ and $HMC_\epsilon$ are averaged over $100$ realizations of each linear reservoir. 
}
\label{fig:6}
\end{figure*}

\begin{figure*}[t]
\centering
\includegraphics[width=16cm,bb=0 0 506 172]{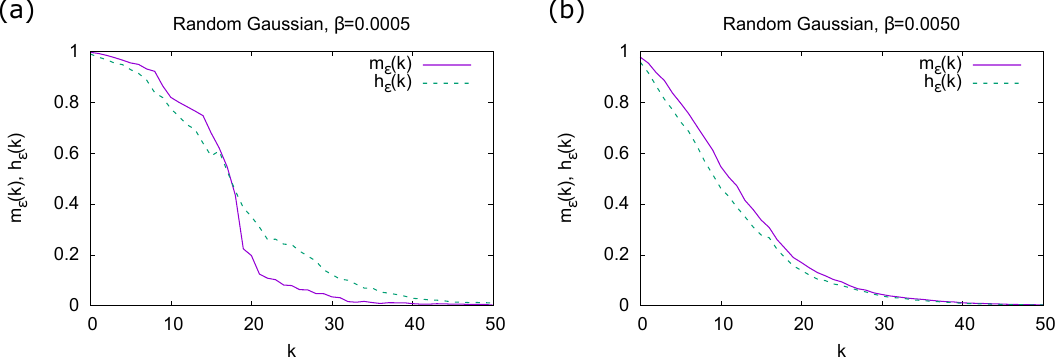}
\caption{
$m_\epsilon(k)$ and $h_\epsilon(k)$ for a single realization of an RGN. (a) $\beta=5.0 \times 10^{-4}$. (b) $\beta=5.0 \times 10^{-3}$. The other parameters are the same as in Fig.~\ref{fig:5}(c).
}
\label{fig:7}
\end{figure*}

\begin{figure*}[t]
\centering
\includegraphics[width=16cm,bb=0 0 511 174]{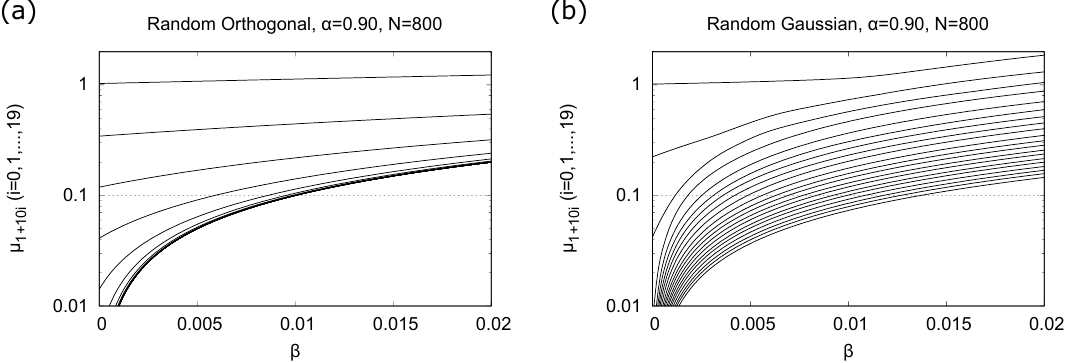}
\caption{
Dependence of the eigenvalues of $C$ on $\beta$ for (a) a RON (b) an RGN. $20$ eigenvalues $\mu_{1+10i}$ ($i=0,1,\dots,19$) are shown. The eigenvalues are obtained from a single realization of a RON and an RGN in Figs.~\ref{fig:5}(b) and \ref{fig:5}(c), respectively. 
}
\label{fig:8}
\end{figure*}

The third example involves random Gaussian networks (RGNs). Let $R=(r_{mn})_{1 \leq m,n \leq N}$ be an $N \times N$ matrix such that each $r_{mn}$ is generated independently following the Gaussian distribution with mean $0$ and variance $1/N$. We take $W=\frac{\sqrt{\alpha}}{\rho}R$, where $\rho$ is the spectral radius of $R$. It seems that $m(k)>h(k)$ holds for $k=0,1,2,\dots$ even as $N \to \infty$ when the noise strength $\beta$ is positive (Figs.~\ref{fig:3} and \ref{fig:4}). One reason for the strict inequality would be that, in contrast to DSRNs and RONs, $C_n$ does not have common eigenvectors with $C_0$ in RGNs. Thus, the noise widens the support of $\bm{\pi}_k$, making it difficult for the equality condition to hold.

Fourth and finally, we take $\bm{f}$ as a sigmoid function such as a hyperbolic tangent function and $L=\mathcal{O}(1)$. The mean-field calculations in the literature imply that $C$ approaches $\mu I$ for some $\mu > 0$~\cite{Haruna2019}. Thus, the equality in Eq.~(\ref{eq:mkhk}) holds asymptotically as $N \to \infty$. Namely, 
\begin{equation}
m(k)=\frac{\| \bm{p}_k \|^2}{\mu}=h(k)
\label{eq:nlinmkhk}
\end{equation}
for $k=0,1,2,\dots$. 

\subsection{Principal Component Regularization of the Reservoir State-Space and Noise-Induced Memory}
\label{subsec:nim}
Next, we investigate the behaviors of $m(k)$ and $h(k)$ under a state-space regularization to elucidate the relation between short-term memory and fluctuations of the reservoir state from a different angle. In particular, we introduce the regularized versions of $m(k)$ and $h(k)$. We observe that the inequality Eq.~(\ref{eq:mkhk}) does not hold for them in general and explain this fact in terms of how the direction of $\overline{\bm{p}}_k$ on which the optimization is performed for the harmonic memory is related to the statistical structure of fluctuations of the whole reservoir state-space. Here, we consider the regularization based on a dimensional reduction of the reservoir state-space~\cite{Lokse2017}, and  we call it \textit{principal component regularization (PCR)}. 

First, we consider the regularization of $m(k)$. Let $K_\epsilon$ be the largest $i$ such that $\mu_i > \epsilon$ for a given nonnegative number $\epsilon < 1$. We optimize $\bm{u} \in \mathbb{R}^L$ after projecting $\bm{x}(t)$ onto the subspace of $\mathbb{R}^L$ spanned by $\bm{q}_1,\bm{q}_2,\dots,\bm{q}_{K_\epsilon}$. The memory function in this case is given by 
\begin{equation}
m_\epsilon(k)=\bm{p}_k^\mathrm{T}D_\epsilon \bm{p}_k,
\label{eq:mke}
\end{equation}
where 
\begin{equation}
D_\epsilon=\sum_{i=1}^{K_\epsilon} \mu_i^{-1}\bm{q}_i \bm{q}_i^\mathrm{T}. 
\label{eq:de}
\end{equation}
The optimal readout weight is given by $\bm{u}=D_\epsilon \bm{p}_k$. 

For $h(k)$, since the optimization of the readout weight is performed with projected states onto $\overline{\bm{p}}_k$, it is natural to consider 
\begin{equation}
h_\epsilon(k) =
\begin{cases}
h(k) & \text{if~~} \overline{\bm{p}}_k^\mathrm{T} C \overline{\bm{p}}_k > \epsilon, \\
0 & \text{otherwise.}
\end{cases}
\label{eq:hke}
\end{equation}

Now we revisit the above three linear networks and study the effect of PCR when the noise strength $\beta$ is varied. In particular, we examine the behaviors of the regularized memory capacity 
\begin{equation}
MC_\epsilon = \sum_{k=0}^\infty m_\epsilon(k)
\label{eq:mce}
\end{equation}
and the regularized harmonic memory capacity 
\begin{equation}
HMC_\epsilon = \sum_{k=0}^\infty h_\epsilon(k). 
\label{eq:hce}
\end{equation}
When $\epsilon=0$, $MC_0$ is just the conventional memory capacity. By Eq.~(\ref{eq:mkhk}), we have $MC_0 \geq HMC_0$ since $m_0(k)=m(k)$ and $h_0(k)=h(k)$ for $k=0,1,2,\dots$. 

For DSRNs, it is straightforward to obtain 
\begin{equation}
MC_\epsilon = \sum_{k=0}^{K_\epsilon - 1} \frac{\alpha^k}{\alpha^k + \tilde{\beta}(1-\alpha^{k+1})} = HMC_\epsilon. 
\label{eq:mcehcedsrn}
\end{equation}
Figure~\ref{fig:5}(a) shows $MC_\epsilon$ $(= HMC_\epsilon)$ for the DSRN with $N=800$, $\alpha=0.9$ and $\epsilon=0.1$. The result for $\alpha=0.8$ is shown in Fig.~\ref{fig:6}(a). $MC_\epsilon$ takes its maximum value at a positive noise strength under PCR. We call this phenomenon \textit{noise-induced memory (NIM)}, and its mechanism can be understood intuitively as follows. Increasing the noise strength $\beta$ degrades the memory of $s(t-k)$ stored in the reservoir state. However, at the same time, it increases the value of $\mu_n$ through $\tilde{\beta}(1-\alpha^n)$. Thus, as $\beta$ is increased continuously from $\beta=0$, the number of $\mu_n$s that exceeds $\epsilon$ increases one by one, which leads to discrete jumps of $MC_\epsilon$ for $\beta < (1-\alpha)\epsilon$ in Figs.~\ref{fig:5}(a) and \ref{fig:6}(a). For $\beta > (1-\alpha)\epsilon$, $MC_\epsilon$ decreases monotonically since the latter effect no longer works: $K_\epsilon = N$. The increment of $MC_\epsilon$ from $\beta=0$ to $\beta=(1-\alpha)\epsilon$ can be estimated as $\frac{\alpha}{\ln \alpha^{-1}}\epsilon \ln \epsilon^{-1}$ upto the leading order of $\epsilon$ by evaluating Eq.~(\ref{eq:mcehcedsrn}) by integrals from both below and above. 

For RONs, we have 
\begin{equation}
HMC_\epsilon = \sum_{k=0}^{K'_\epsilon} \frac{\alpha^k}{\alpha^k + \tilde{\beta}}
\label{eq:hceron}
\end{equation}
by Eqs.~(\ref{eq:ronmkhk}) and (\ref{eq:hke}) under the annealed approximation, where $K'_\epsilon$ is the largest $k$ such that $\alpha^k + \tilde{\beta} > \epsilon$. It seems that $MC_\epsilon$ is also approximated well by Eq.~(\ref{eq:hceron}) for $N \to \infty$ [Figs.~\ref{fig:5}(b) and \ref{fig:6}(b)]. This can be explained as follows. $\tilde{\bm{v}}_k$ and $\tilde{\bm{v}}_l$ ($k \neq l$) are almost orthogonal in the sense that their inner product is $\mathcal{O}(\frac{1}{\sqrt{N}})$ with high probability~\cite{Vershynin2018}. If we assume that $\tilde{\bm{v}}_k^\mathrm{T} \tilde{\bm{v}}_l = \mathcal{O}(\frac{1}{\sqrt{N}})$ for $0 \leq k \neq l \leq K'_\epsilon$ and $K'_\epsilon$ is a sufficiently large number independent of $N$, then we can argue that $\mu_n$ is close to $\alpha^{n-1} + \tilde{\beta}$ and $\tilde{\bm{v}}_{n-1}$ approximately satisfies the condition to be an eigenvector of $C$ corresponding to $\mu_n$ for $n=1,2,\dots,K_\epsilon$ for $N \to \infty$ (Appendix~\ref{sec:a3}). Independent of this argument, the increment of $\mu_n$ due to the noise is $\tilde{\beta}$. Thus, the mechanism of NIM explained above works as in DSRNs, and Eq.~(\ref{eq:hceron}) has a peak at $\beta = (1-\alpha)\epsilon$ [Figs.~\ref{fig:5}(b) and \ref{fig:6}(b)]. The increment of Eq.~(\ref{eq:hceron}) from $\beta=0$ to $\beta=(1-\alpha)\epsilon$ can be approximated as $\frac{\epsilon}{\ln \alpha^{-1}}$ for $\epsilon \ll 1$. 

For RGNs, the increment of $\mu_n$ due to the noise cannot be calculated separately from the value of $\mu_n$ when $\beta=0$, and there seems to be no simple formula for it. We still observe NIM numerically, although its effect is weaker than that in DSRNs and RONs [Figs.~\ref{fig:5}(c) and \ref{fig:6}(c)]. The relation between $MC_\epsilon$ and $HMC_\epsilon$ in this case has a remarkable feature: $MC_\epsilon < HMC_\epsilon$ for sufficiently small $\beta$. For this to occur, there must be a $k$ such that $m_\epsilon(k) < h_\epsilon(k)$. This inequality implies that $h_\epsilon(k)$ retains a part of $m(k)-m_\epsilon(k)$, the part of $m(k)$ ignored under PCR. Indeed, let us consider the decomposition $h_\epsilon(k) - m_\epsilon(k) = \left( h_\epsilon(k) - m(k) \right) + \left( m(k) - m_\epsilon(k) \right)$. Since $h_\epsilon(k) \leq h(k) \leq m(k)$, we have $h_\epsilon(k) - m(k) \leq 0$. For $h_\epsilon(k) - m_\epsilon(k) > 0$ to hold, $m(k) - m_\epsilon(k) = \sum_{\mu_i \leq \epsilon} m_i(k)\pi_{k,i}$ must be positive, namely, $\pi_{k,i} = \left( \overline{\bm{p}}_k^\mathrm{T}\bm{q}_i \right)^2 > 0$ for some $i$ with $\mu_i \leq \epsilon$. This means that the direction of $\overline{\bm{p}}_k$ on which the amount of memory $h_\epsilon(k)$ is stored overlaps the subspace spanned by $\bm{q}_i$ with $\mu_i \leq \epsilon$. Figure~\ref{fig:7} shows $m_\epsilon(k)$ and $h_\epsilon(k)$ numerically obtained from a single realization of an RGN with (a) $\beta=5.0 \times 10^{-4}$ and (b) $\beta=5.0 \times 10^{-3}$. $m_\epsilon(k) < h_\epsilon(k)$ holds for $k \geq 18$ for $\beta=5.0 \times 10^{-4}$, while there is no such $k$ for $\beta=5.0 \times 10^{-3}$. 

We note that NIM is a cooperative effect of the two regularizations, PCR and the addition of noise. Each regularization itself degrades the reservoir's short-term memory capability. On one hand, if the reservoir is not subject to noise, the application of PCR decreases memory functions in general by ignoring the part of input signals stored in the direction of eigenvectors corresponding to the eigenvalues $\mu_i \leq \epsilon$. On the other hand, if PCR is not introduced, then only the memory-degrading effect of the noise exists. In particular, NIM does not occur if we fix the number of $\mu_i$s that are used in Eq.~(\ref{eq:de}) instead of fixing the threshold $\epsilon$ as in PCR, since the increase of $MC_\epsilon$ is caused by the increase of $K_\epsilon$. However, the increase of $K_\epsilon$ is due to the increase of the noise strength $\beta$, which in turn decreases $\mu_i^{-1}$ in Eq.~(\ref{eq:de}). Thus, the competition between these two opposing effects resulting from the combination of PCR and the addition of noise is at the heart of NIM. The presence of NIM implies that the former effect is dominant up to a certain value of $\beta$. However, the degree of dominance of the former effect depends on the reservoir network architectures as shown in Figs.~\ref{fig:5} and \ref{fig:6}. Figure~\ref{fig:8} compares dependence of the eigenvalues $\mu_i$ of $C$ on the noise strength $\beta$ between RONs and RGNs in Fig.~\ref{fig:5}. In the case of RONs, $MC_\epsilon$ continues to increase upto $\beta=(1-\alpha)\epsilon=0.01$ where all $\mu_i$s exceed $\epsilon=0.1$ [Figs.~\ref{fig:5}(b) and \ref{fig:8}(a)]. In contrast, in the case of RGNs, $MC_\epsilon$ begins to decrease around $\beta=0.01$ [Fig.~\ref{fig:5}(c)]. However, there are still many $\mu_i$s below $\epsilon=0.1$ at $\beta=0.01$ and they exceed $\epsilon$ at larger values of $\beta$ [Fig.~\ref{fig:8}(b)]. The directions in the state-space corresponding to these $\mu_i$s do not overcome the memory-degrading effect of the noise. Thus, NIM is not just an artifact of PCR, but reflect the intrinsic fluctuation structure of the reservoir states. 

\section{Concluding Remarks}
\label{sec:conclusion}
The mechanism of NIM is reminiscent of recurrence resonance~\cite{Krauss2019,Metzner2024}, in which the mutual information between the current state of the system and that of the next time step is maximized at a certain strength of noise. The noise degrades temporal correlation in the state of the system on one hand, and it enlarges and uniformizes the region of the state-space visited by the system on the other hand. The balance between these two competing effects leads to a peak of the mutual information as the noise strength is varied. Similarly, two competing effects are involved in NIM as explained in the previous section: the noise added to the input signals blurs the memory of them in the current state on one hand, and the number of available directions in the state-space to store the input signals increases on the other hand. 

\begin{figure}[t]
\centering
\includegraphics[width=8cm,bb=0 0 339 234]{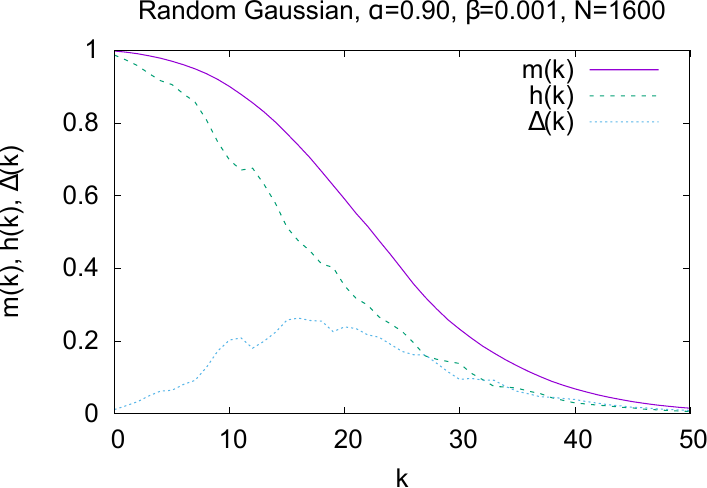}
\caption{
$m(k)$ and $h(k)$ for a single realization of an RGN with $\alpha=0.9$, $\beta=0.001$ and $N=1600$. $h(11) < h(12)$ and $\Delta(12) < \Delta(11)$ holds. Note that $\Delta(k+1) < \Delta(k)$ does not necessarily imply $h(k) < h(k+1)$ as can be seen from most of $k \geq 16$. 
}
\label{fig:9}
\end{figure}

The memory functions $m(k)$ for linear ESNs with $L=N$ are known to be monotonically decreasing functions of $k$~\cite{Jaeger2002}. Namely, $m(k) \geq m(k+1)$ holds for $k=0,1,2,\dots$ for a single realization of such ESN. However, $h(k)$ does not necessarily decrease monotonically. Indeed, we have 
\begin{equation}
h(k) - h(k+1) \geq \Delta(k+1) - \Delta(k). 
\label{eq:nondec}
\end{equation}
If $\Delta(k+1) < \Delta(k)$, then the inverted inequality $h(k) < h(k+1)$ is allowed (Fig.~\ref{fig:9}). We note that if $h(k) < h(k+1)$ holds, then we have the strict inequality $m(k) > h(k)$. This argument is formally similar to the justification of consistency between the second law of thermodynamics and Maxwell's demon in general information exchange processes~\cite{Sagawa2012,Sagawa2013}, in that the inequality that holds in the whole system does not necessarily restrict to subsystems. This could indicate a common mathematical framework subsuming both of them, which is not pursued further here. 

\begin{figure*}[t]
\centering
\includegraphics[width=16cm,bb=0 0 509 173]{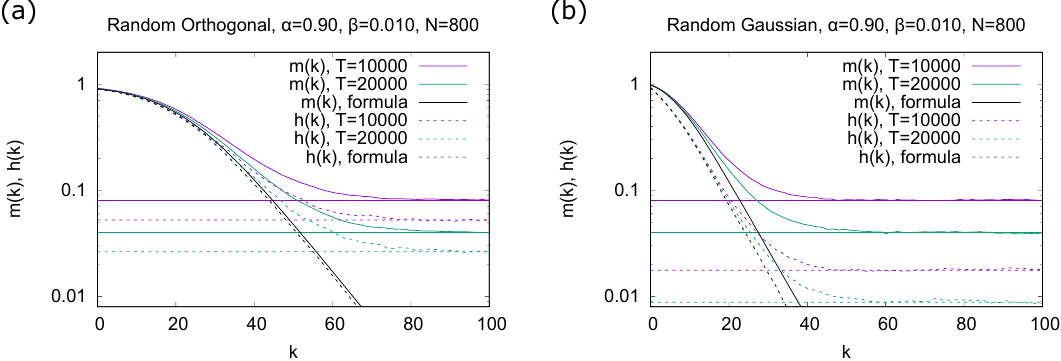}
\caption{
$m(k)$ (solid lines) and $h(k)$ (broken lines) estimated by numerical simulations of Eq.~(\ref{eq:esn}) over finite time lengths $T=10000, 20000$ for (a) RONs and (b) RGNs with $N=800$, $\alpha=0.9$, and $\beta = 0.01$. Both quantities are averaged over $20$ trials with random initial conditions. In each trial, the initial $1000$ time steps are ignored. The values of $m(k)$ and $h(k)$ calculated from the theoretical formulas of $\bm{p}_k$ and $C$ [Eqs.~(\ref{eq:pklin}) and (\ref{eq:clin}), respectively] are also shown. Horizontal lines are the theoretical predictions of the overestimates of $m(k)$ (solid lines) and $h(k)$ (broken lines) when they are close to zero (The upper and lower lines are for $T=10000$ and $T=20000$, respectively). They are $\frac{N}{T}$ and $c\frac{N}{T}$ with $c = \frac{\left(\operatorname{tr}C\right)^2}{N\operatorname{tr}\left(C^2\right)}$, respectively, where $C$ is given by Eq.~(\ref{eq:clin}). 
}
\label{fig:10}
\end{figure*}

In this paper, we focus mainly on theoretical and conceptual issues about the harmonic memory function $h(k)$ and its relation to $m(k)$. In particular, in linear cases we rely on the formulas for $\bm{p}_k$ and $C$ (Eqs.~(\ref{eq:pklin}) and (\ref{eq:clin}), respectively) to compute $m(k)$ and $h(k)$ without actually simulating Eq.~(\ref{eq:esn}). It is known that there is an overestimate of order $\mathcal{O}\left( \frac{L}{T} \right)$ for $m(k) \approx 0$ if we estimate it by numerical simulations over a finite time length $T$~\cite{Dambre2012}. Figure~\ref{fig:10} confirms this fact for RONs and RGNs. Since $h(k) \leq m(k)$, we expect that the overestimate of $h(k)$ when $m(k) \approx 0$ is smaller than that of $m(k)$ in general. Indeed, we can show that the average overestimate of $m(k)$ is just $\frac{L}{T}$ and that for $h(k)$ is $c\frac{L}{T}$ with $c = \frac{\left(\operatorname{tr}C\right)^2}{L\operatorname{tr}\left(C^2\right)} \leq 1$ for large $L$ and $T$ with $L \ll T$ under certain assumptions modeling the situation $m(k)=0$ (Appendix~\ref{sec:a4}). In Figs.~\ref{fig:10}(a) and \ref{fig:10}(b) where $L=N=800$, $c \approx 0.658$ and $c \approx 0.220$, respectively. Another source of numerical inaccuracy for $m(k)$ is the evaluation of the small eigenvalues of $C$. One requires careful numerical treatment when evaluating $m(k)$ due to these two sources of numerical inaccuracy~\cite{Ballarin2024}. For $h(k)$, the second issue is mitigated since it does not depend on $C^{-1}$. 

Recently, the usefulness of the memory capacity as a measure of inherent information processing capability of reservoir network architectures has been questioned since it achieves its maximum value for generic reservoir networks in linear ESNs without input noise, and it can vary depending on the scale of input signals in nonlinear ESNs~\cite{Ballarin2025}. Here, we give a few comments on this criticism. First, our approach in this paper focuses on the performance of ESNs relative to a specific task, namely, the short-term memory capability. Second, we have studied the short-term memory capability of ESNs in terms of the fluctuation structure of their state-space and how input noise modifies it. Ballarin et al.~\cite{Ballarin2025} lacks this viewpoint. Third, our interest is not only on the memory capacity itself (called the total memory capacity in Ballarin et al.~\cite{Ballarin2025}), but also on the memory functions (called the $\tau$-lag memory capacity in Ballarin et al.~\cite{Ballarin2025}) which have richer information on the relation between the short-term memory capability and the fluctuation structure of the state-space. For example, we have explained the reason why the inequality $MC_\epsilon < HMC_\epsilon$ holds for sufficiently small noise strength $\beta$ in RGNs by examining the difference between $m_\epsilon(k)$ and $h_\epsilon(k)$ in how they utilize the state-space to store the input signals. Fourth and finally, we note that if we regard the scale of the input signals as a control parameter, the fact that the memory capacity can reflect it in nonlinear systems can be an advantage for exploring an effective way to use nonautonomous dynamical systems as reservoirs in the context of physical reservior computing where the architecture of reservoirs themselves could be difficult to adjust, as we discussed in our previous work~\cite{Haruna2019}. 

In conclusion, we have presented an inequality representing an uncertainty relation associated with the short-term memory capability of ESNs. It gives a lower bound for the memory function. The lower bound itself can be regarded as a suboptimal memory and is called the harmonic memory function. We found that the uncertainty relation does not necessarily hold for the regularized memory and harmonic memory functions under the state-space regularization called PCR and this fact is related to NIM induced by PCR. The inequality holds as long as both sides of Eq.~(\ref{eq:mkhk}) can be defined, and thus has broad applicability in a wide range of dynamical systems. 

\appendix

\section{Derivation of Eq.~(\ref{eq:deltak})}
\label{sec:a1}
First, we show that $\hat{y'}_k(t)$ is orthogonal to $\hat{y}_k(t) - \hat{y'}_k(t)$, namely, 
\begin{equation}
\left\langle \hat{y'}_k(t) \left( \hat{y}_k(t) - \hat{y'}_k(t) \right) \right\rangle_t = 0, 
\label{eqsm:ip}
\end{equation}
where $\hat{y}_k(t)$ and $\hat{y'}_k(t)$ are scalar variables defined by Eqs.~(\ref{eq:hatyk}) and (\ref{eq:hatypk}), respectively. Indeed, 
\begin{align}
&\left\langle \hat{y'}_k(t) \left( \hat{y}_k(t) - \hat{y'}_k(t) \right) \right\rangle_t \nonumber\\
&= \sigma_k \left\langle \bm{p}_k^\mathrm{T} \bm{x}(t) \left( (C^{-1} - \sigma_k I) \bm{p}_k \right)^\mathrm{T} \bm{x}(t) \right\rangle_t \nonumber\\
&= \sigma_k \bm{p}_k^\mathrm{T} \left\langle \bm{x}(t) \bm{x}(t)^\mathrm{T} \right\rangle_t (C^{-1} - \sigma_k I) \bm{p}_k \nonumber\\
&= \sigma_k \bm{p}_k^\mathrm{T}C(C^{-1} - \sigma_k I) \bm{p}_k \nonumber\\
&= \sigma_k \bm{p}_k^\mathrm{T}(I - \sigma_k C) \bm{p}_k \nonumber\\
&= \sigma_k \left( \| \bm{p}_k \|^2 - \sigma_k \bm{p}_k^\mathrm{T} C \bm{p}_k \right) \nonumber\\
&= \sigma_k \left( \| \bm{p}_k \|^2 - \| \bm{p}_k \|^2 \right) = 0, 
\label{eqsm:ipder}
\end{align}
where we used Eqs.~(\ref{eq:hatyk}) and (\ref{eq:hatypk}) in the first equality, Eq.~(\ref{eq:c}) in the third equality, and Eq.~(\ref{eq:sigk}) in the sixth equality. 

Second, we express $\hat{y}_k(t) - \hat{y'}_k(t)$ as 
\begin{align}
\hat{y}_k(t) - \hat{y'}_k(t) &= \bm{p}_k^\mathrm{T} \left( C^{-1} - \sigma_k I \right) \bm{x}(t) \nonumber\\
&= \bm{p}_k^\mathrm{T} \left( \sum_{i=1}^L \left( \mu_i^{-1} - \sigma_k \right) \bm{q}_i \bm{q}_i^\mathrm{T} \right) \bm{x}(t) \nonumber\\
&= \sum_{i=1}^L \left( \mu_i^{-1} - \sigma_k \right) \bm{p}_k^\mathrm{T} \bm{q}_i \bm{q}_i^\mathrm{T} \bm{x}(t), 
\label{eqsm:ydiff}
\end{align}
where we used Eqs.~(\ref{eq:hatyk}) and (\ref{eq:hatypk}) in the first equality and Eq.~(\ref{eq:c-1}) in the second equality. 

Now, we have 
\begin{align}
\Delta(k) &= \langle \hat{y}_k(t)^2 \rangle_t - \langle \hat{y'}_k(t)^2 \rangle_t \nonumber\\
&= \left\langle \left( \hat{y}_k(t) - \hat{y'}_k(t) \right)^2 \right\rangle_t \nonumber\\
&= \left\langle \left( \sum_{i=1}^L \left( \mu_i^{-1} - \sigma_k \right) \bm{p}_k^\mathrm{T} \bm{q}_i \bm{q}_i^\mathrm{T} \bm{x}(t) \right)^2 \right\rangle_t \nonumber\\
&= \sum_{i,j=1}^L \left( \mu_i^{-1} - \sigma_k \right) \left( \mu_j^{-1} - \sigma_k \right) \nonumber\\
&\qquad \qquad \times \bm{p}_k^\mathrm{T} \bm{q}_i \bm{q}_i^\mathrm{T} \left\langle \bm{x}(t) \bm{x}(t)^\mathrm{T} \right\rangle_t \bm{q}_j \bm{q}_j^\mathrm{T} \bm{p}_k \nonumber\\
&= \sum_{i,j=1}^L \left( \mu_i^{-1} - \sigma_k \right) \left( \mu_j^{-1} - \sigma_k \right) \bm{p}_k^\mathrm{T} \bm{q}_i \bm{q}_i^\mathrm{T} C \bm{q}_j \bm{q}_j^\mathrm{T} \bm{p}_k \nonumber\\
&= \sum_{i=1}^L \left( \mu_i^{-1} - \sigma_k \right)^2 \mu_i \left( \bm{p}_k^\mathrm{T} \bm{q}_i \right)^2 \nonumber\\
&= \sum_{i=1}^L \left( m_i(k) - h(k) \right)^2 \frac{\pi_{k,i}}{m_i(k)} \nonumber\\
&= \frac{1}{h(k)}\sum_{i=1}^L \left( m_i(k) - h(k) \right)^2 \rho_{k,i}, 
\label{eqsm:deltak}
\end{align}
where the second equality follows from Eq.~(\ref{eqsm:ip}) and we used Eq.~(\ref{eqsm:ydiff}) in the third equality, Eq.~(\ref{eq:c}) in the fifth equality, Eq.~(\ref{eq:cexpansion}) in the sixth equality, Eqs.~(\ref{eq:hk}), (\ref{eq:piki}), (\ref{eq:mik}), and (\ref{eq:sigk}) in the seventh equality, and Eq.~(\ref{eq:rhoki}) in the eighth equality. 

We note that Eq.~(\ref{eq:deltak}) can also be derived from an expression of the difference between the arithmetic mean $A = \sum_{j=1}^J a_j r_j$ and the harmonic mean $H = \frac{1}{\sum_{j=1}^J \frac{r_j}{a_j}}$ of positive real numbers $a_1,\dots,a_J$ weighted by a probability distribution $(r_1,\dots,r_J)$: 
\begin{align}
A - H &= \sum_{j=1}^J \left( a_j - H \right) r_j \nonumber\\
&= \sum_{j=1}^J \left( a_j - H \right) \left( 1 - \frac{H}{a_j} \right) r_j 
= \sum_{j=1}^J \left( a_j - H \right)^2  \frac{r_j}{a_j}. 
\label{eqsm:ah}
\end{align}

\section{Derivation of Eq.~(\ref{eq:ronmkhk})}
\label{sec:a2}
Fix $k \geq 0$. Let 
\begin{equation}
C_0 = C - \alpha^k Q^k \bm{v}\bm{v}^\mathrm{T}Q^{-k}. 
\label{eq:c0ron}
\end{equation}
We have 
\begin{equation}
\bm{p}_k^\mathrm{T} C \bm{p}_k = \alpha^{2k} + \alpha^k r, 
\label{eq:pkcpkron}
\end{equation}
where we used Eqs.~(\ref{eq:pkron}), (\ref{eq:cron}) and (\ref{eq:tildevk}), and we put 
\begin{equation}
r = \bm{v}^\mathrm{T}Q^{-k} C_0 Q^k \bm{v}. 
\label{eq:r}
\end{equation}
Let $\bm{v}_1, \dots, \bm{v}_N$ be an orthonormal basis of $\mathbb{R}^N$ such that 
\begin{equation}
\bm{v}_1 = Q^k \bm{v} 
\label{eq:v1}
\end{equation}
and $\bm{v}_2, \dots, \bm{v}_N$ are chosen uniformly at random in the orthogonal subspace of $\bm{v}_1$. Since $\sum_{n=1}^N \bm{v}_n \bm{v}_n^\mathrm{T} = I$, we have $\operatorname{tr}C_0 = \operatorname{tr}\left( C_0 \sum_{n=1}^N \bm{v}_n \bm{v}_n^\mathrm{T} \right) = \sum_{n=1}^N \operatorname{tr} \left( C_0 \bm{v}_n \bm{v}_n^\mathrm{T} \right)$. Let $\mathrm{E}\left[ \cdot \right]$ denote the average over the choice of $Q$, $\bm{v}$ and $\bm{v}_2, \dots, \bm{v}_N$. Then, $\mathrm{E} \left[ \operatorname{tr} \left( C_0 \bm{v}_n \bm{v}_n^\mathrm{T} \right) \right]$ should not depend on $n$ since $\bm{v}_1$ is independent of $C_0$ under the assumption of the annealed approximation. Thus, we have 
\begin{align}
\mathrm{E}\left[ r \right] 
&= \mathrm{E}\left[ \bm{v}_1^\mathrm{T} C_0 \bm{v}_1 \right] \nonumber\\
&= \mathrm{E}\left[ \operatorname{tr} \left( C_0 \bm{v}_1 \bm{v}_1^\mathrm{T} \right) \right] \nonumber\\
&= \frac{1}{N} \mathrm{E}\left[ \operatorname{tr} C_0 \right], 
\label{eq:er0}
\end{align}
where we used Eqs.~(\ref{eq:r}) and (\ref{eq:v1}) in the first equality. Now, 
\begin{align}
\operatorname{tr} C_0 
&= \operatorname{tr}\left( C - \alpha^k Q^k \bm{v}\bm{v}^\mathrm{T}Q^{-k} \right) \nonumber\\
&= \operatorname{tr}C - \alpha^k \operatorname{tr}\left( Q^k \bm{v}\bm{v}^\mathrm{T}Q^{-k} \right) \nonumber\\
&= \operatorname{tr}\left( \sum_{l=0}^\infty \alpha^l \tilde{\bm{v}}_l \tilde{\bm{v}}_l^\mathrm{T} + \tilde{\beta}I \right) - \alpha^k \operatorname{tr}\left( \tilde{\bm{v}}_k \tilde{\bm{v}}_k^\mathrm{T} \right) \nonumber\\
&= \sum_{l=0}^\infty \alpha^l \operatorname{tr}\left( \tilde{\bm{v}}_l \tilde{\bm{v}}_l^\mathrm{T} \right) + \tilde{\beta}\operatorname{tr}I - \alpha^k \operatorname{tr}\left( \tilde{\bm{v}}_k \tilde{\bm{v}}_k^\mathrm{T} \right) \nonumber\\
&= \sum_{k=0}^\infty \alpha^k + \tilde{\beta}N - \alpha^k \nonumber\\
&= \frac{1}{1-\alpha} + \tilde{\beta}N - \alpha^k, 
\label{eq:trc0}
\end{align}
where we used Eq.~(\ref{eq:c0ron}) in the first equality, Eqs.~(\ref{eq:cron}) and (\ref{eq:tildevk}) in the third equality, and the fifth equality holds since $\operatorname{tr}\left( \tilde{\bm{v}}_k \tilde{\bm{v}}_k^\mathrm{T} \right) = \operatorname{tr}\left( \tilde{\bm{v}}_k^\mathrm{T} \tilde{\bm{v}}_k \right) = \operatorname{tr}1 = 1$ and $\operatorname{tr}I = N$. From Eqs.~(\ref{eq:er0}) and (\ref{eq:trc0}), we find 
\begin{equation}
\mathrm{E}\left[ r \right] = \tilde{\beta} + \mathcal{O}\left( \frac{1}{N} \right). 
\label{eq:er}
\end{equation}

Finally, we find 
\begin{align}
\mathrm{E}\left[ h(k) \right] 
&= \mathrm{E}\left[ \frac{\| \bm{p}_k \|^2}{\overline{\bm{p}}_k^\mathrm{T}C\overline{\bm{p}}_k} \right] \nonumber\\
&= \mathrm{E}\left[ \frac{\alpha^k}{\alpha^k + r }\right] \nonumber\\
&\geq \frac{\alpha^k}{\alpha^k + \mathrm{E}\left[ r \right] } \to \frac{\alpha^k}{\alpha^k + \tilde{\beta}} = m(k), 
\label{eqsm:hk}
\end{align}
as $N \to \infty$, where we used Eqs.~(\ref{eq:pkron}) and (\ref{eq:pkcpkron}) in the second equality, and Jensen's inequality and Eq.~(\ref{eq:er}) in the third line.

\section{Eigenvalues and eigenvectors of $C$ for RONs in the context of NIM}
\label{sec:a3}
Let $K'_\epsilon$ be the largest $k$ such that $\alpha^k + \tilde{\beta} > \epsilon$ as in the main text. We assume that $K'_\epsilon$ is a constant independent of $N$. We put $\nu_{kl} = \tilde{\bm{v}}_k^\mathrm{T}\tilde{\bm{v}}_l$. We assume that $\nu_{kl} = \mathcal{O}(\frac{1}{\sqrt{N}})$ for $0 \leq k \neq l \leq K'_\epsilon$. 

Let 
\begin{equation}
\bm{d}_k = C \tilde{\bm{v}}_k - \left( \alpha^k + \tilde{\beta} \right) \tilde{\bm{v}}_k. 
\end{equation}
Since $C$ is given by Eq.~(\ref{eq:cron}) for RONs, we have 
\begin{equation}
\bm{d}_k = \sum_{l \neq k} \alpha^l \nu_{kl} \tilde{\bm{v}}_l. 
\label{eq:eigenron}
\end{equation}

For $k \leq K'_\epsilon$, we evaluate 
\begin{align}
\left\| \bm{d}_k \right\|^2 
&= \sum_{l,l' \neq k} \alpha^{l+l'} \nu_{kl} \nu_{kl'} \nu_{ll'} \nonumber\\
&= \sum_{k \neq l,l' \leq K'_\epsilon} \alpha^{l+l'} \nu_{kl} \nu_{kl'} \nu_{ll'} \nonumber\\
&\qquad + \sum_{l,l' > K'_\epsilon} \alpha^{l+l'} \nu_{kl} \nu_{kl'} \nu_{ll'}. 
\label{eq:norm2eigenron}
\end{align}
The first sum in Eq.~(\ref{eq:norm2eigenron}) is $\mathcal{O}(\frac{1}{N})$ since $\nu_{kl} \nu_{kl'} = \mathcal{O}(\frac{1}{N})$ and the sum has a finite number of terms independent of $N$. For the second sum in Eq.~(\ref{eq:norm2eigenron}), we have 
\begin{equation}
\left\| \sum_{l,l' > K'_\epsilon} \alpha^{l+l'} \nu_{kl} \nu_{kl'} \nu_{ll'} \right\|
\leq \alpha^{2K'_\epsilon + 2} \sum_{l,l' \geq 0} \alpha^{l}\alpha^{l'} = \left(\frac{\alpha^{K'_\epsilon + 1}}{1 - \alpha} \right)^2. 
\end{equation}
Thus, Eq.~(\ref{eq:norm2eigenron}) can be arbitrarily small as $N \to \infty$ and $K'_\epsilon$ is sufficiently large. In particular, there exists $g_{N,K'_\epsilon} > 0$ such that 
\begin{equation}
\left\| \bm{d}_k \right\| < g_{N,K'_\epsilon} 
\label{eq:g}
\end{equation}
for all $0 \leq k \leq K'_\epsilon$ and $ \lim_{K'_\epsilon \to \infty} \lim_{N \to \infty} g_{N,K'_\epsilon} = 0$. 
 
Next, we show that there is no eigenvalue $\mu_n$ of $C$ that cannot be approximated by the form $\alpha^k + \tilde{\beta}$ for $1 \leq n \leq K_\epsilon$. To show this claim, we appeal to a proof by contradiction. Let $\mu$ be an eigenvalue of $C$ such that $\mu > \epsilon$, namely, $\mu = \mu_n$ for some $1 \leq n \leq K_\epsilon$, and assume that $\mu$ cannot be approximated by any of $\alpha^k + \tilde{\beta}$ with $0 \leq k \leq K'_\epsilon$. Let $\tilde{\bm{v}}$ be a normalized eigenvector corresponding to $\mu$. Since $\mu$ is different from $\alpha^k + \tilde{\beta}$ for all $0 \leq k \leq K'_\epsilon$, there exists $\xi > 0$ such that 
\begin{equation}
\min_{0 \leq k \leq K'_\epsilon} \left\{ \left| \alpha^k + \tilde{\beta} - \mu \right| \right\} > \xi.
\label{eq:xi}
\end{equation}
Note that $\xi$ does not depend on $K'_\epsilon$ if $K'_\epsilon$ is sufficiently large. 

We have 
\begin{align}
\mu &= \tilde{\bm{v}}^\mathrm{T}C\tilde{\bm{v}} \nonumber\\
&= \sum_{k=0}^\infty \alpha^k \left( \tilde{\bm{v}}^\mathrm{T}\tilde{\bm{v}}_k \right)^2 + \tilde{\beta} = S_1 + S_2 + \tilde{\beta},
\label{eq:mu}
\end{align}
where 
\begin{equation}
S_1 = \sum_{k=0}^{K'_\epsilon} \alpha^k \left( \tilde{\bm{v}}^\mathrm{T}\tilde{\bm{v}}_k \right)^2
\label{eq:s1}
\end{equation}
and 
\begin{equation}
S_2 = \sum_{k = K'_\epsilon + 1}^\infty \alpha^k \left( \tilde{\bm{v}}^\mathrm{T}\tilde{\bm{v}}_k \right)^2 \leq \sum_{k = K'_\epsilon + 1}^\infty \alpha^k = \frac{\alpha^{K'_\epsilon + 1}}{1-\alpha}. 
\label{eq:s2}
\end{equation}
$S_2$ can be arbitrarily small if $K'_\epsilon$ is sufficiently large. For $S_1$, since $C$ is symmetric, we have 
\begin{align}
\mu \tilde{\bm{v}_k}^T \tilde{\bm{v}} 
&= \tilde{\bm{v}_k}^T C \tilde{\bm{v}} \nonumber\\
&= \left( C\tilde{\bm{v}_k} \right)^T \tilde{\bm{v}} \nonumber\\
&= \left( \alpha^k + \tilde{\beta} \right) \tilde{\bm{v}_k}^T \tilde{\bm{v}} + \bm{d}_k^T \tilde{\bm{v}}. 
\label{eq:vkv}
\end{align}
From Eqs.~(\ref{eq:vkv}), (\ref{eq:g}) and (\ref{eq:xi}), 
\begin{align}
\left| \tilde{\bm{v}_k}^T \tilde{\bm{v}} \right| 
&= \frac{\left| \bm{d}_k^T \tilde{\bm{v}}  \right|}{\left| \alpha^k + \tilde{\beta} - \mu \right|} \nonumber\\
& < \frac{g_{N,K'_\epsilon}}{\xi}
\label{eq:vkvabs}
\end{align}
for all $0 \leq k \leq K'_\epsilon$. Thus, we obtain 
\begin{equation}
S_1 = \sum_{k=0}^{K'_\epsilon} \alpha^k \left( \tilde{\bm{v}}^\mathrm{T}\tilde{\bm{v}}_k \right)^2 < \frac{g_{N,K'_\epsilon}^2}{(1-\alpha)\xi^2} 
\label{eq:s1ineq}
\end{equation}
by Eqs.~(\ref{eq:s1}) and (\ref{eq:vkvabs}). It follows that $S_1 \to 0$ as $N \to \infty$ and $K'_\epsilon \to \infty$. From Eqs.~(\ref{eq:mu}), (\ref{eq:s2}) and (\ref{eq:s1ineq}), 
\begin{equation}
\mu \to \tilde{\beta} \leq \epsilon - \alpha^{K'_\epsilon + 1} < \epsilon 
\label{eq:muepsilon}
\end{equation}
as $N \to \infty$ and $K'_\epsilon \to \infty$, where the first inequality holds since $\alpha^{K'_\epsilon + 1} + \tilde{\beta} \leq \epsilon$ by the definition of $K'_\epsilon$. However, Eq.~(\ref{eq:muepsilon}) contradicts the assumption $\mu > \epsilon$. 

In summary, up to the approximation error, $\mu_n = \alpha^{n-1} + \tilde{\beta}$ and $\tilde{\bm{v}}_{n-1}$ satisfies the condition to be an eigenvector of $C$ corresponding to $\mu_n$ for $n=1,2,\dots, K_\epsilon$.

\section{$m(k)$ and $h(k)$ under finite time average}
\label{sec:a4}
In this appendix, we study the overestimates of $m(k)$ and $h(k)$ when they are close to zero if the infinite time average $\langle \cdot \rangle_t$ is replaced by a finite time average $\langle \cdot \rangle_{t,T} = \frac{1}{T}\sum_{t=0}^{T-1}(\cdot)$. We model the situation $m(k)=h(k)=0$ as follows. Let $\nu > \mu_1 \geq \mu_2 \geq \dots \geq \mu_L > \lambda > 0$ with $\nu, \lambda$ constants independent of $L$, and $\bm{q}_1,\bm{q}_2,\dots,\bm{q}_L$ unit vectors in $\mathbb{R}^L$ that form an orthonormal basis of it. We assume that $\bm{x}(t)=\sum_{i=1}^L a_i(t)\bm{q}_i$ where $a_i(t)$ are mutually independent stochastic variables with mean $0$ and variance $\mu_i$ for $i=1,2,\dots,L$ and $t=0,1,2,\dots$. We also assume that each $a_i(t)$ is independent of input signals $s(t')$ for $t'=0,1,2,\dots$ and all the moments of $a_i(t)$ and $s(t)$ are finite. In this appendix, $\mathrm{E}\left[ \cdot \right]$ denote the average over $a_i(t)$ and $s(t)$. 

Let $C_T = \langle \bm{x}(t)\bm{x}(t)^\mathrm{T} \rangle_{t,T}$ and $\bm{p}_{k,T} = \langle s(t-k)\bm{x}(t) \rangle_{t,T}$. It is straightforward to see that 
\begin{equation}
\mathrm{E}\left[ C_T \right] = \sum_{i=1}^L \mu_i \bm{q}_i \bm{q}_i^\mathrm{T}.
\end{equation}
Thus, $\mu_1,\mu_2,\dots,\mu_L$ are the eigenvalues of $\mathrm{E}\left[ C_T \right]$ and $\bm{q}_1,\bm{q}_2,\dots,\bm{q}_L$ are corresponding eigenvectors. $\lambda, \nu$ are lower and upper bounds of the eigenvalues that are not dependent on $L$, respectively. These exist for the three linear models in the main text with positive noise strength $\beta > 0$. In the following, we find that $\mathrm{E}\left[ m(k) \right] = \frac{L}{T}$ and $\mathrm{E}\left[ h(k) \right] = \mathcal{O}\left(\frac{L}{T}\right)$. Since one can show that the deviations of $C_T$ from $\mathrm{E}\left[ C_T \right]$ add higher order terms of $\frac{L}{T}$ to $\mathrm{E}\left[ m(k) \right]$ and $\mathrm{E}\left[ h(k) \right]$ after some algebra, we ignore them in the following calculations by assuming $L \ll T$, and write $C$ for $\mathrm{E}\left[ C_T \right]$. 

We put $A_i = \langle s(t-k)a_i(t) \rangle_{t,T}$ for $i = 1,2,\dots,L$. Since $\bm{p}_{k,T} = \sum_{i=1}^L A_i \bm{q}_i$, we have 
\begin{equation}
m(k) = \bm{p}_{k,T}^\mathrm{T} C^{-1} \bm{p}_{k,T} = \sum_{i=1}^L \mu_i^{-1} A_i^2, 
\end{equation}
\begin{equation}
\| \bm{p}_{k,T} \|^2 = \sum_{i=1}^L A_i^2, 
\end{equation}
and 
\begin{equation}
\bm{p}_{k,T}^\mathrm{T} C \bm{p}_{k,T} = \sum_{i=1}^L \mu_i A_i^2.
\end{equation}
Using $\mathrm{E}\left[ A_i \right] = 0$ and $\mathrm{E}\left[ A_i^2 \right] = \frac{\mu_i}{T}$, we obtain 
\begin{equation}
\mathrm{E}\left[ m(k) \right] = \frac{1}{T}\sum_{i=1}^L \mu_i^{-1}\mu_i = \frac{L}{T}, 
\end{equation}
\begin{equation}
e_1 = \mathrm{E}\left[ \| \bm{p}_{k,T} \|^2 \right] = \frac{1}{T}\sum_{i=1}^L \mu_i = \frac{\operatorname{tr}C}{T}, 
\end{equation}
and 
\begin{equation}
e_2 = \mathrm{E}\left[ \bm{p}_{k,T}^\mathrm{T} C \bm{p}_{k,T} \right] = \frac{1}{T}\sum_{i=1}^L \mu_i^2 = \frac{\operatorname{tr}\left(C^2\right)}{T}. 
\end{equation}
For $\mathrm{E}\left[ h(k) \right]$, the following inequality holds. 
\begin{equation}
\mathrm{E}\left[ h(k) \right] \geq \frac{e_1^2}{e_2} = c\frac{L}{T},
\label{eq:hkineq}
\end{equation}
where we used the Cauchy-Schwarz inequality for $\sqrt{\bm{p}_{k,T}^\mathrm{T} C \bm{p}_{k,T}}$ and $\frac{\| \bm{p}_{k,T} \|^2}{\sqrt{\bm{p}_{k,T}^\mathrm{T} C \bm{p}_{k,T}}}$, and put 
\begin{equation}
c = \frac{\left(\operatorname{tr}C\right)^2}{L\operatorname{tr}\left(C^2\right)}.
\end{equation}
 We note that $c \leq 1$, which also follows from the Cauchy-Schwarz inequality. 

Next, we quantitatively evaluate the gap between both sides of Eq.~(\ref{eq:hkineq}) for large $L$ and $T$ with $L \ll T$. We put 
\begin{equation}
\gamma_1 = \frac{\| \bm{p}_{k,T} \|^2 - e_1}{e_1}
\end{equation}
and 
\begin{equation}
\gamma_2 = \frac{\bm{p}_{k,T}^\mathrm{T} C \bm{p}_{k,T} - e_2}{e_2}. 
\end{equation}
We have 
\begin{align}
h(k) &= \frac{\left( e_1 + \left( \| \bm{p}_{k,T} \|^2 - e_1 \right) \right)^2}{e_2 + \left( \bm{p}_{k,T}^\mathrm{T} C \bm{p}_{k,T} - e_2 \right)} \nonumber\\
&= \frac{e_1^2}{e_2} \frac{(1+\gamma_1)^2}{1 + \gamma_2} \nonumber\\
&= \frac{e_1^2}{e_2} (1+\gamma_1)^2 (1 - \gamma_2 + \gamma_2^2 - \dots) \nonumber\\
&= \frac{e_1^2}{e_2} (1 + 2\gamma_1 - \gamma_2 + \gamma_1^2 - 2\gamma_1 \gamma_2 + \gamma_2^2 - \dots) \nonumber\\
&= \frac{e_1^2}{e_2} \left(1 + 2\gamma_1 - \gamma_2 + (\gamma_1 - \gamma_2)^2 - \dots\right). 
\label{eq:hkexpansion}
\end{align}
The third line in Eq.~(\ref{eq:hkexpansion}) can be justified by $\mathrm{E}\left[ \gamma_2 \right] = 0$ and $\mathrm{Var}\left( \gamma_2 \right) = \mathrm{E}\left[ \gamma_2^2 \right] = \mathcal{O}\left( \frac{1}{L} \right)$, where $\mathrm{Var}\left( X \right)$ is the variance of a given stochastic variable $X$. It also holds that $\mathrm{E}\left[ \gamma_1 \right] = 0$ and $\mathrm{Var}\left( \gamma_1 \right) =\mathrm{E}\left[ \gamma_1^2 \right]  = \mathcal{O}\left( \frac{1}{L} \right)$. To obtain these order estimates of $\mathrm{E}\left[ \gamma_1^2 \right]$ and $\mathrm{E}\left[ \gamma_2^2 \right]$, we need $\mathrm{Var}\left( A_i^2 \right)$. Since $\mathrm{E}\left[ A_i^4 \right] = \frac{3 \mu_i^2}{T^2} + \mathcal{O}\left( \frac{1}{T^3} \right)$ after some algebra, we have 
\begin{equation}
\mathrm{Var}\left( A_i^2 \right) 
= \mathrm{E}\left[ \left( A_i^2 \right)^2 \right] - \mathrm{E}\left[ A_i^2 \right]^2 
= \frac{2\mu_i^2}{T^2} + \mathcal{O}\left( \frac{1}{T^3} \right). 
\end{equation}
Thus, 
\begin{align}
\mathrm{E}\left[ \gamma_1^2 \right] 
&= \frac{\sum_{i=1}^L \mathrm{Var}\left( A_i^2 \right)}{e_1^2} \nonumber\\
&= \frac{2\operatorname{tr}\left( C^2 \right) + \mathcal{O}\left( \frac{L}{T} \right)}{\left( \operatorname{tr}C \right)^2} = \mathcal{O}\left( \frac{1}{L} \right)
\label{eq:eg12}
\end{align}
and similarly 
\begin{equation}
\mathrm{E}\left[ \gamma_2^2 \right] = \frac{2\operatorname{tr}\left( C^4 \right) + \mathcal{O}\left( \frac{L}{T} \right)}{\left( \operatorname{tr}\left( C^2 \right) \right)^2} = \mathcal{O}\left( \frac{1}{L} \right), 
\label{eq:eg22}
\end{equation}
where we used $0 < \lambda \leq \mu_i < \nu$ for $i=1,2,\dots,L$ in the last equalities in Eqs.~(\ref{eq:eg12}) and (\ref{eq:eg22}). 

Now, we take the average of the right-hand side of Eq.~(\ref{eq:hkexpansion}) as follows. 
\begin{align}
\mathrm{E}\left[ h(k) \right] &= \frac{e_1^2}{e_2} \left( 1 + 2\mathrm{E}\left[ \gamma_1 \right] - \mathrm{E}\left[ \gamma_2 \right] + \mathrm{E}\left[ (\gamma_1 - \gamma_2)^2 \right] - \dots \right) \nonumber\\
&= c\frac{L}{T} \left( 1 + \mathrm{E}\left[ (\gamma_1 - \gamma_2)^2 \right] - \dots \right). 
\end{align}
To calculate $\mathrm{E}\left[ (\gamma_1 - \gamma_2)^2 \right]$, we need 
\begin{align}
\mathrm{E}\left[ \gamma_1 \gamma_2 \right] = \frac{1}{e_1 e_2}\left( \sum_{i,j=1}^L \mu_i \mathrm{E}\left[ A_i^2 A_j^2 \right] - e_1 e_2 \right). 
\label{eq:eg1g2_0}
\end{align}
Since 
\begin{equation}
\mathrm{E}\left[ A_i^2 A_j^2 \right] = \frac{\mu_i \mu_j}{T^2} + \mathcal{O}\left( \frac{1}{T^3} \right)
\end{equation}
for $i \neq j$, we have 
\begin{align}
\sum_{i,j=1}^L \mu_i \mathrm{E}\left[ A_i^2 A_j^2 \right] &= \sum_{i=1}^L \mathrm{E}\left[ A_i^4 \right] + \sum_{i \neq j} \mu_i \mathrm{E}\left[ A_i^2 A_j^2 \right] \nonumber\\
&= \frac{3 \sum_{i=1}^L \mu_i^3}{T^2} + \frac{\sum_{i \neq j} \mu_i^2 \mu_j}{T^2} + \mathcal{O}\left( \frac{L^2}{T^3} \right) \nonumber\\
&= \frac{2 \sum_{i=1}^L \mu_i^3}{T^2} + \frac{\sum_{i,j=1}^L \mu_i^2 \mu_j}{T^2} + \mathcal{O}\left( \frac{L^2}{T^3} \right) \nonumber\\
&= \frac{2 \operatorname{tr}\left( C^3 \right)}{T^2} + e_1 e_2 + \mathcal{O}\left( \frac{L^2}{T^3} \right). 
\label{eq:eg1g2_1}
\end{align}
Substituting Eq.~(\ref{eq:eg1g2_1}) into Eq.~(\ref{eq:eg1g2_0}), 
\begin{align}
\mathrm{E}\left[ \gamma_1 \gamma_2 \right] &= \frac{1}{e_1 e_2}\left( \frac{2 \operatorname{tr}\left( C^3 \right)}{T^2} + \mathcal{O}\left( \frac{L^2}{T^3} \right) \right) \nonumber\\
&= \frac{2 \operatorname{tr}\left( C^3 \right)}{\operatorname{tr}C \operatorname{tr}\left( C^2 \right)} + \mathcal{O}\left( \frac{1}{T} \right). 
\label{eq:eg1g2}
\end{align}
Combining Eqs.~(\ref{eq:eg12}), (\ref{eq:eg22}) and (\ref{eq:eg1g2}), we obtain 
\begin{equation}
\mathrm{E}\left[ (\gamma_1 - \gamma_2)^2 \right] = \eta + \mathcal{O}\left( \frac{1}{T} \right), 
\label{eq:diffg1g2}
\end{equation}
where 
\begin{equation}
\eta = 2 \left( \frac{\operatorname{tr}\left( C^2 \right)}{\left( \operatorname{tr}C \right)^2} + \frac{\operatorname{tr}\left( C^4 \right)}{\left( \operatorname{tr}\left( C^2 \right) \right)^2} - 2 \frac{\operatorname{tr}\left( C^3 \right)}{\operatorname{tr}C \operatorname{tr}\left( C^2 \right)} \right). 
\end{equation}
Note that $\eta \geq 0$ by the Cauchy-Schwarz inequality. 

In summary, the ratio of $\mathrm{E}\left[ h(k) \right]$ to its lower bound in Eq.~(\ref{eq:hkineq}) is 
\begin{equation}
\frac{\mathrm{E}\left[ h(k) \right]}{c\frac{L}{T}} = 1 + \eta, 
\end{equation}
if we ignore the $\mathcal{O}\left( \frac{1}{T} \right)$ term in Eq.~(\ref{eq:diffg1g2}). Since $\eta = \mathcal{O}\left( \frac{1}{L} \right)$, we expect that the equality in Eq.~(\ref{eq:hkineq}) approximately holds for large $T$ and $L$ with $L \ll T$. For example, in Figs.~\ref{fig:10}(a) and \ref{fig:10}(b), $\eta \approx 0.0283$ and $\eta \approx 0.0205$, respectively. 

Finally, we note that if $\eta = 0$ and $T \to \infty$ then $\mathrm{E}\left[ (\gamma_1 - \gamma_2)^2 \right] = 0$ by Eq.~(\ref{eq:diffg1g2}). This in turn implies that $\gamma_1 = \gamma_2$ almost everywhere, which is the equality condition of Eq.~(\ref{eq:hkineq}).

\begin{acknowledgments}
TH was supported by JSPS KAKENHI Grants (No.~JP23K03219 and No.~JP23K17485). TH and KN were supported by the SECOM Science and Technology Foundation. 
\end{acknowledgments}

\section*{Data availability statements}
The data that support the findings of this article are openly available~\cite{DAS}. 

%

\end{document}